# A New Mechanism For Mutual Authentication In *SIP*

[1]Maisam Mohammadian, [2]Naser Mozayani

**Abstract**
The greatest threat in the new generation network which is called NGN is unsafe authentication. Communication between new servers in NGN world is done based on Session Initiation Protocol (SIP). SIP is an application-layer control operating on top of a transport protocol which allows creating, modifying, and terminating sessions among more agents. For authentication, SIP relies on HTTP Digest by default; the client is authenticated to the SIP proxy server, called one way authentication, because in this approach we can authenticate client to server and the client can't do any authentication in server side. In this paper we propose a mutual authentication mechanism that is not based on HTTP Digest and then, we implement our method in IMS and start to do authentication client to server is done in first step and server to client next. By this solution, we found that our proposed mechanism has more stability against intruder and unauthorized access to system. In this paper, we introduce a new approach for user authentication or server authentication by using OPTIONAL SIP's header. In our solution we can increase necessary steps for intruder to decode authentication, compilation time and encoding time and save our system from some attack such as offline attack and replay

## 1. Introduction

The primary generation of networks services was based on client server architectures. The modern networks like IMS, considers the new need for security and implement some mutual authentication mechanisms: for example Authentication and Key Agreement (AKA) [1] and TLS and IPSec [2] are respectively deployed for mobile networks to mutually authenticate the entities using challenge-response mechanisms. In this paper, we provide a new approach to increase authentication security between client and SIP servers. SIP is a real time signaling protocol used in IMS networks. Usually, SIP is used in many applications such as chat, voice over IP (VOIP), any real time services, etc.
SIP uses a set of request messages including INVITE, ACK, CANCEL, OPTIONS, BYE and REGISTER. A User Agent, wishing to initiate a session, sends an INVITE message. SIP uses OK message after ACK message and uses BYE to terminate session. More explanation can be found in the RFC3261 document [3]. Some solutions have been developed to improve SIP security, but most of protocols or implementations create interoperability problems. Attackers can easily download free software to sniff packets and analyzes networks [4]. IP telephony infrastructure could be attacked by using a large collection of hack techniques [5].Since the authentication based on HTTP Digest mechanism is one of SIP weakness, this paper proposes a new approach to improve authentication in SIP security base on IMS network.

## 2. SIP authentication mechanism

Using SIP, there exist three basic principles for providing security, Tasi, Yang and Durlanik [6]. These three security solutions on SIP data involves http digest. In terms of authentication in SIP, several mechanisms exist which we introduce as follows. SIP specification [8] introduces HTTP digest authentication and usage of S/MIME extensions. In addition to those, we introduce also other techniques of authentication, which are designed to improve security in SIP. HTTP Digest provides a challenge/response message and is used by default MD5 algorithm [9]. As another aspect, some lightweight programs such as lightweight scheme defined by Kong et al have several problems. For example, this mechanism uses SIP user client phones in signing contact addresses instead of the registrar servers. The actual amount of throughput is not given in [6].
SIP uses a header called 'nonce' to build a 'response' in the general protocol that use at now. The nonce is a random variable which is generated by server proxy and is created using username, method and password for response to operate. The response produced by this formula is shown in algorithm(1) as shown in fig. 1.

- H(A1)=MD5(username: realm: password)
- H(A2)=MD5(METHOD: Request-URI)
- response=MD5(H(A1): nonce: H(A2))

**Figure 1:** Algorithm1**:** a creating response

## 3. Problems in SIP authentication

We have two major weaknesses in HTTP digest authentication in SIP [4]. The first missing security issue is the lack of securing all headers and parameters in SIP which would possibly need protection. The second security weakness, relating to digest authentication, is the requirement of pre-existing user configuration on servers, which does not scale well [9].Though, for authenticating in cellular mobile communication, it provides simple authentication. With the Universal Mobile Telecommunications System, there is an improvement with the AKA mechanism [9]. This solution enables a mutual authentication between any devices and the network. This security policy requires a shared secret key and a shared cryptographic algorithm that exist in SIP. So, pre-share keys are one of the main problems for security distributed keys and cause algorithm load for encode and decode security information packet. S/MIME in SIP is used on carrying signed or encrypted replication of headers and authenticating users. This mechanism lacks the public key distribution problem, which means that the public keys used in authentication are difficult to distribute and maintain. The public key infrastructure is also susceptible to man-in-the middle attack [10,11]. It uses several hash computations and server certificates to ensure security. This causes overhead and reduction in performance. So, we tried to solve these problems without any protocol changes. Also, if the load increases, the server comes more vulnerable to denial of service attacks as stated in [12, 13].

## 4. OUR SOLUTION OVERVIEW

First, when invite message is received in server, the server start to generate nonce and qop and give a copy of these two parameters and then these two values are encrypted by MD5. Then, it sends these 2 parameters by its IP and "nc" to client side which we use this massage in challenge massage. In the client side, decoding this packet and saving "qop", "nonce" and server's IP and changing client "nc" to one. After received "qop", the client start to generate "cnonce" and make copy and save of it for compare in next step. Cnonce a new sip parameter that introduced in RFC 2617 this parameter is random value generated after receive qop in server challenge message after generated cnonce by client. we show this algorithm in Fig 2.

> Authenticate: Digest
> realm="biloxi.com",
> qop="auth,auth-int",
> Server ip="X.X.X.X",
> nonce="dcd98b7102dd2f0e8b11d0f600bfb0",
> nc= for first try is one
>
> algorithm=MD5

**Figure 2:** Algorithm1: generating nonce and qop and nc in server proxy

Second, after the execution of the first part, the random value generated is called cnonce. Then, we apply SIP messages and encrypt them by MD5 algorithm. The other side in the server proxy uses the algorithm (3) to save cnonce and compare (qop, nc, nonce) which is shown in Fig. 3 . The Cnonce is calculated in the client by this algorithm (3).

> Authorization: Digest username="",
> Realm="biloxi.com",
> nonce="dcd98b7102dd2f0e8b11d0f600bfb0",
> qop=auth,
> nc=00000001,
> IP="X.X.X.X",
> cnonce="0a4f113b",
> response="6629faea05397450978507c4ef1",

**Figure 3: Algorithm2:** Cnonce genareate in client and send to server

Finally, when message is received in the proxy server, the proxy starts to decode the message and compare the qop and nonce messages to save qop, nonce and, nc. The related algorithm is shown in Fig. 3 and the flowchart is demonstrated in Fig. 4. If these three parameters are the same, then the user is authenticated for the server and server sends Cnonce, qop, nonce and nc adds on 200 ok massage. However, if one of these items is missing and it is found through comparing, the client will not be authenticated and the session will be terminated. These two algorithms are necessary for mutual authentication since we should use three parameters that are useful for mutual authentication. At the client side, when the 200 ok message is received, the add-on parameters start to compare the nonce, cnonce, qop and nc parameters. If all parameters are same as the previous message, the server is authenticated to the client. This model is suitable for service providers and users because at the service side, providers do not need to change any protocols in subsystem and, its extra lightweight character is suitable for weaker client's devices. By applying this approach we will be able to access the selective mutual authentication. The selective mutual authentication invitation massage is shown in algorithm (4). In other words; the algorithm (4) is our only suggestion in case of selective algorithm in user's device. This selective announce will be the first sent message to the server after the invite message. Nevertheless, if the client chooses the selective algorithm in his mutual authentication, he will send a second message to the server Another approach can also be implemented in which client can selectively authenticate server. If it is not necessary to authenticate, by sending the following algorithm it does not authenticate server. In proposed algorithm, if client does not send value of cnonce, server just send message 200 ok and decrease client overhead.

Client sends authentication to server, algorithm of request-digest when cnonce is not presented and client does not want to authenticate server and the MD5 algorithm is used, it is shown in below.

This concept means that the client sends the algorithm (4) to server when authenticating the server to the client is not necessary, he sends the algorithm (2) when it is necessary to authenticate server. We tested this idea in mutual authentication by AVSPA tools.

```
request-digest = <">< KD ( H(A1), unq(nonce-value)
 ":" nc-value
  ":"unq(qop-value)
 ":" H(A2)
   ) ><">
A1 = unq(username-value) ":" unq(realm-value) ":" passwd
A2 = Method ":" digest-uri-value
```

**Figure :** Algorithm4: not Cnonce genareateand client does not need to server authentication

Our simulations are done using AVSPA tools. We compared our results with Yang, Durlanik, Tasi. We also found actual amount of throughput and compared it by Tasi,Yang, Durlanik and two other solution [16,10]. Our results are shown in the Table1.

For more clarification we test our approach in real HTTP Digest that worked in traditional and we found several results which are shown in table below. Their approach increased the processing overhead of SIP messages by 20 ms in trade. The initial overhead was 50 which rose to 70 ms with their method. Considering analysis, their approach caused the stages to reach 5 which were 4. As another aspect of our method, we can reduce our processing time. We compare this method with Yang, Durlanik, Tasi. We increase steps of intruder for accesses to session line from 26 to 32 steeps using this solution.

The comparison is summarized in Table 1

Table 1: Margin specifications

| Parameter | Traditional Digest | New approach | Yang | Durlanik | Tasi |
|---|---|---|---|---|---|
| Intruder | 26 step | **32 step** | 30step | 31step | 30step |
| Compilation | 0.07ms | **0.02ms** | 0.03ms | 0.06ms | 0.04ms |
| Encoder | 0.02 ms | **0.01ms** | 0.04ms | 0.04 | 0.03 |
| Analyze | 5 stag | **4 stage** | 4 | 5 | 4 |
| Overhead | 70ms | **50ms** | 80ms | 85ms | 80ms |
| Throughput | 30ms | **20ms** | 20ms | 30ms | 20ms |

This solution performs approximately near to using SIP networks. We have tested encoding time, processing time, intruder steeps and throughput of a SIP server for authentication in call setup phase. We also evaluate the results by the amount of INVITE requests served per minute. The considerable reduction in processing and encoding time without reducing network security is novelty of our method in NGN network. This method has very lightweight scheme and it is suitable for mutual authentication in SIP. It is not dependent on structure. This method can effectively solve problems about offline attacks.

In our solution we can solve problems regarding distributed keys by our method by generating new string of new value.

## Conclusion

### Acknowledgments

Authors want to thank from he Iranian Telecommunication Researcch center for ICT – ITRC for supporting this research work at IMS innovation Lab at ITRC-Iran-Tehran.

### References

[1] J. Rosenberg, H. Schulzrinne, G. Camarillo, A. Johnston, J. Peterson, R. Sparks, M. Handley and E. Schooler,"SIP: Session Initiation Protocol", RFC 3261, June 2002: http://www.ietf.org/rfc/rfc3261.txt..

[2] B. Wallis, "Hypertext Transfer Protocol (HTTP) Digest Authentication using Global System for Mobile Communications (GSM) A3 and A8", Internet draft (work in progress), August 2007.

[3] Liao YP, Wang SS. "A new secure password authenticated key agreement scheme for SIP using self-certified public keys on elliptic curves". *Computer and Communications* 2010; **33**(3):372–380.

[4] P. Urien, "TLS Tripartite Diffie-Hellman Key Exchange", IETF draft (work in progress), July 2010.

[5] RFC 4492, "Elliptic Curve Cryptography (ECC) Cipher Suites for Transport Layer Security (TLS)", May 2006.

[6] T. Russell, The IP Multimedia Subsystem (IMS) Session Control &

Other Network Operations, published by Mc Graw Hill,2008.

[7] W. Werapun, A. A. El Kalam, B. Paillassa, J. Fasson," Solution Analysis for SIP Security Threats", The IEEE international Conference on Innovations In Inforamtion Technology(IIT'08), 2008.

[8] B. Li,D. Wang, S. Zhang, " Policy Based SIP Signaling Management in IMS", Alcatel-Lucent 1-6/F Sec. D Hi-Tech Software Park Qingdao, 266101, China

[9] A. Nemi, J. Arkko, V. Torvinen, " Hypertext Transfer Protocol(HTTP) Digest Authentication Using Authentication and Key Agrement(AKA),IETF RFC 3310, 2002.

[10] T. Dierks, E. Rescorla, The Transport |Layer Security(TLS) Protocol Version 1.2, 2008

[11] M. Sher, T. Magedanz, "Developing Network Domain Security (NDS) Model for IP Multimedia Subsystem (IMS)" ,IEEE First International Conference on Availability, Reliability and Security (ARES'06), Austria, 2006

[12] F. Leitold, A. Medve, L. Kovacs, " SIP security problems in NGM Services ",  The IEEE International Conference on Next Generation Mobile Applications, Services and Technologies (NGMAST'07),2007.

[13] www.avspa.com

[14] Jo H, Lee Y, Kim M, Kim S, Won D. "Off-line password-guessing attack to Yang's and Huang's authentication schemes for session initiation protocol". *Fifth International Joint Conference on INC, IMS and IDC*, 2009; 618–621.

[15] . Tsai JL. "Efficient nonce-based authentication scheme for session initiation protocol". *International Journal of Network Security* 2009; **8**(3):312–316.

[16] . Yoon E, Shin Y, Jeon I, Yoo K. "Robust mutual authentication with a key agreement scheme for the session initiation protocol". *IETE Technical Review* 2010; **27**(3):203–213.

[17] Kong L., Balasubramaniyan V.B. and Ahamad M. A lightweight scheme for securely and reliably locating SIP users. VoIP Management and Security, IEEE Workshop, Apr 2006. Page(s): 9-17.
.

**First Author** 2010-2012: M.Sc. student in Computer Architecture, Computer Engineering Department, Iran University of Science and Technology, Tehran, Iran. GPA:

2007-2009: B.Sc. In Information Communication Technology ,

Electrical  Engineering Department, Kermanshah. University Of

Applied Science And Technology. IRAN, Tehran. GPA:

.

**Second Author** Ph.D.  Informatics, University of Rennes I, FRANCE, July 1998. M.Sc. Telematics & Information Systems, SUPELEC, Rennes, FRANCE, June 1994 B.Sc. Electrical Engineering (Computer Hardware), Sharif University ofTechnology, Tehran, IRAN, Sept,1990
.